# From the up-converting multimodal luminescent thermometer to ratiometric visual power density meter based on $Er^{3+},Yb^{3+}$ emission

**Anam Javaid[1], Maja Szymczak[1*], Lukasz Marciniak[1*]**

[1] Institute of Low Temperature and Structure Research, Polish Academy of Sciences,

Okólna 2, 50-422 Wrocław, Poland

Corresponding authors: l.marciniak@intibs.pl

m.szymczak@intibs.pl

**Abstract**

This study demonstrates that thermally induced variations in the spectroscopic properties of $Na_3Sc_2(PO_4)_3$:$Er^{3+}$,$Yb^{3+}$ can be effectively harnessed for multimodal remote temperature sensing. As shown, $Na_3Sc_2(PO_4)_3$:$Er^{3+}$,$Yb^{3+}$ supports multiple ratiometric sensing modes based on the intensity ratios of (i) $^2H_{11/2} \rightarrow {}^4I_{15/2}$ and $^4S_{3/2} \rightarrow {}^4I_{15/2}$; (ii) $^2H_{9/2} \rightarrow {}^4I_{13/2}$ and $^4S_{3/2} \rightarrow {}^4I_{15/2}$; and (iii) green-to-red emission intensity ratio, achieving maximum relative sensitivities of 2.8%K$^{-1}$, 3% K$^{-1}$, and 1.8% K$^{-1}$, respectively. The synergy between thermal changes observed in the green-to-red emission intensity ratio of $Er^{3+}$ ions, combined with the efficient optical heating of $Na_3Sc_2(PO_4)_3$:$Er^{3+}$,$Yb^{3+}$ at elevated $Yb^{3+}$ concentrations enables the development of a visual optical power density sensor, exhibiting relative sensitivities of $S_{Rx}$ = 1.0 % W$^{-1}$ cm$^2$ and $S_{Ry}$ = 0.9 % W$^{-1}$ cm$^2$ at 15 W cm$^{-2}$ when quantified using CIE 1931 chromaticity coordinates. To the best of our knowledge, this is the first report of a visual luminescent optical power density sensor. Furthermore, it was demonstrated that $Na_3Sc_2(PO_4)_3$:$Er^{3+}$,$Yb^{3+}$ can be successfully applied for two-dimensional imaging of optical power density, thereby enabling spatial visualization of power distribution within an illuminated field.

**Introduction**

Up-converting phosphors co-doped with $Er^{3+}$ and $Yb^{3+}$ ions have garnered significant attention in recent years, primarily due to their potential applications in remote temperature sensing[1–10]. This capability arises from the thermal coupling between the $^2H_{11/2}$ and $^4S_{3/2}$ energy levels of $Er^{3+}$ ions[11,12]. According to the Boltzmann distribution, an increase in temperature enhances the relative population of the $^2H_{11/2}$ level compared to the $^4S_{3/2}$ level[11,13]. The resulting temperature-dependent luminescence intensity ratio (*LIR*) between emissions from these levels offers a theoretically predictable thermometric parameter, representing a key advantage for quantitative temperature readout. However, the close spectral overlap of the $^2H_{11/2} \rightarrow {}^4I_{15/2}$ and $^4S_{3/2} \rightarrow {}^4I_{15/2}$ transitions necessitates the use of bandpass filters for effective thermal imaging.

In this study, we present results for $Na_3Sc_2(PO_4)_3$:$Er^{3+}$,$Yb^{3+}$, a material in which, alongside the typical $Er^{3+}$-related $^2H_{11/2} \rightarrow {}^4I_{15/2}$, $^4S_{3/2} \rightarrow {}^4I_{15/2}$, and $^4F_{3/2} \rightarrow {}^4I_{15/2}$ emission bands, additional luminescence corresponding to the $^2H_{9/2} \rightarrow {}^4I_{13/2}$ transition was detected. This emission results from a three-photon excitation process, enabling population of the $^2H_{9/2}$ state. The unique population dynamics in $Na_3Sc_2(PO_4)_3$:$Er^{3+}$,$Yb^{3+}$ support three distinct temperature readout strategies - *LIR* between: (i) $^2H_{11/2} \rightarrow {}^4I_{15/2}$ and $^4S_{3/2} \rightarrow {}^4I_{15/2}$; (ii) $^2H_{9/2} \rightarrow {}^4I_{13/2}$ and $^4S_{3/2} \rightarrow {}^4I_{15/2}$; and (iii) green-to-red emission intensity ratio. These approaches provide application-specific flexibility in thermometric performance. Notably, the third method enables direct visual temperature indication, as increasing temperature enhances green up-conversion emission with respect to red one, producing a perceptible thermal color shift in the emitted light. Furthermore, the 980 nm excitation light absorbed by $Yb^{3+}$ ions induces optical heating of the $Na_3Sc_2(PO_4)_3$:$Er^{3+}$,$Yb^{3+}$ material, with efficiency increasing with sensitizer ($Yb^{3+}$) concentration. Leveraging this effect, we introduce the first luminescence-based visual optical power density meter. As demonstrated, a standard digital camera and analysis of green and red channel signals, enables visualization of the spatial distribution of 980 nm beam intensity without additional optical filters, providing simple, rapid and filter-free approach to optical power density mapping.

**Experimental section**

*Synthesis*

Powder samples of $Na_3Sc_2(PO_4)_3$:Er,0.5%,x%Yb$^{3+}$, (where x = 5, 10, 15, 30) were synthesized using a conventional high-temperature solid-state reaction technique. $Na_2CO_3$ (99.9% of purity, Alfa Aesar), $Sc_2O_3$ (99.9% of purity, Alfa Aesar), $NH_4H_2PO_4$ (99.9% of purity, POL-AURA), $Er_2O_3$ (99.999% of purity, Stanford Materials Corporation) and $Yb_2O_3$ (99.999% of purity, Stanford Materials Corporation) were used as starting materials. The stoichiometric amounts of reagents were finely ground in an agate mortar with few drops of hexane and, then annealed in the alumina crucibles at 1573 K for 5 h (heating rate of 10 K min$^{-1}$) in air. The final powders were allowed to cool naturally to the room temperature and then ground again to obtain powder samples for structural and optical characterization.

*Characterization*

The obtained materials were examined using powder X-ray diffraction technique. Powder diffraction data were obtained in Bragg–Brentano geometry using a PANalytical X'Pert Pro diffractometer using Ni-filtered Cu K$\alpha$ radiation (V=40 kV, I=30 mA).

A differential scanning calorimetric (DSC) measurements were performed using Perkin-Elmer DSC 8000 calorimeter equipped with Controlled Liquid Nitrogen Accessory LN2 with a heating/cooling rate of 20 K min$^{-1}$. The sample was sealed in the aluminum pan. The measurement was performed for the powder sample in the 100 - 800 K temperature range.

The excitation spectra were obtained using the FLS1000 Fluorescence Spectrometer from Edinburgh Instruments equipped with 450 W Xenon lamp and R928 photomultiplier tube from Hamamatsu as a detector. Emission spectra were measured using the same system with 980 nm laser diodes as excitation source. During the temperature-dependent emission measurements, the temperature of the sample was controlled by a THMS600 heating–cooling stage from Linkam (0.1 K temperature stability and 0.1 K set point resolution). Luminescence decay profiles were also recorded using the FLS1000 equipped with 150 W µFlash lamp.

To obtain the images, photographs were taken using a Canon EOS 400D camera equipped with an EFS 60 mm macro lens and a 750 nm short-pass optical filter (Thorlabs). Luminescence images were acquired under 980 nm laser excitation. The temperature of the samples was determined using a FLIR T540 thermographic camera, providing a measurement accuracy of ±0.5 K. The red and green channels (RGB) were extracted from the photographs using IrfanView 64 4.51 software, and the R and G intensity maps were subsequently processed and divided using ImageJ 1.8.0_172 software.

**Results and discussion**

$Na_3Sc_2(PO_4)_3$ - a member of the NASICON (sodium superionic conductor) family of compounds - is reported to exist in three different crystallographic structures[14–18]. The specific structure in which $Na_3Sc_2(PO_4)_3$ crystallizes depends on synthesis conditions, ionic substitution schemes, and external stimuli such as temperature[14–24]. In general, temperature-induced reversible phase transitions have been widely reported: upon heating, the monoclinic $α$ phase with space group *Bb* transforms into the $β$ phase at approximately 320 K, and subsequently into the $γ$ phase at around 440 K, both of which adopt a trigonal structure with space group $R\bar{3}c$ [16,18]. This behavior was confirmed by differential scanning calorimetry (DSC) analysis performed for $Na_3Sc_2(PO_4)_3$ sample co-doped with 5%$Yb^{3+}$ and 0.5%$Er^{3+}$ ions (Figure S1), where a reversible phase transition was observed at approximately 340 K. However, important factor influencing the structure of $Na_3Sc_2(PO_4)_3$ is ionic substitution. This phenomenon has been demonstrated in the case of $Na_3Sc_2(PO_4)_3$ host ions substitution by lanthanide ions[16,17,23]. For instance, it has been shown that increasing concentrations of $Eu^{2+}$ and $Eu^{3+}$ ions induce a transition from a monoclinic to a trigonal structure[23]. A similar effect is observed in the present study for $Yb^{3+}$ and $Er^{3+}$ co-doping, as evidenced by the XRD patterns measured. For $Yb^{3+}$ concentrations of 5% and 10% (with constant $Er^{3+}$ concentration of 0.5%), a monoclinic phase at room temperature was obtained, while at 15%, a gradual transition toward a trigonal phase occurred, resulting in a mixed monoclinic-trigonal phase composition. In contrast, the sample with 30% $Yb^{3+}$ exhibits a fully trigonal phase. Furthermore, the XRD measurements confirmed that the incorporation of up to 30% $Yb^{3+}$ does not lead to the formation of impurity phases or structural defects. Additionally, no evidence of this phase transition was observed for this sample in the DSC analysis (Figure S2). The difference in the ionic radii between $Yb^{3+}$ ($R \approx$ 0.868 Å) and $Er^{3+}$ ($R \approx$ 0.89 Å) replacing $Sc^{3+}$ ($R \approx$ 0.745 Å)[25], leading to a contraction of the unit cell with an increase in the $Yb^{3+}$ concentration reflected in the shift of the diffraction reflections toward larger 2theta angles.

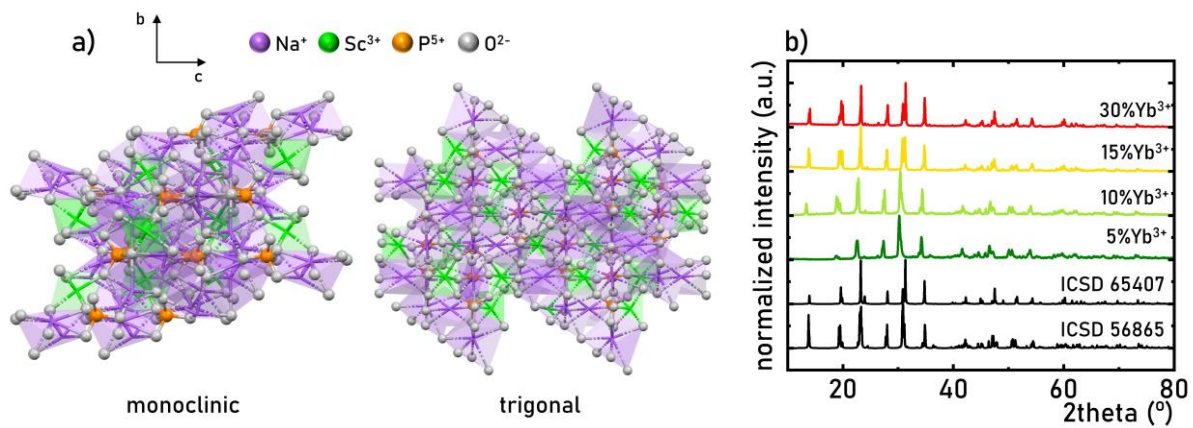

**Figure 1**. Visualization of the structure of the monoclinic and trigonal phases of $Na_3Sc_2(PO_4)_3$-a) the comparison of room temperature XRD patterns of $Na_3Sc_2(PO_4)_3$:$Er^{3+}$, $Yb^{3+}$ with different $Yb^{3+}$ concentration-b).

The up-conversion properties of the $Er^{3+}$, $Yb^{3+}$-doped systems have been extensively studied[12,26–36]. Therefore, only a simplified description is provided here. In such systems, $Yb^{3+}$ ions serve as sensitizers by absorbing excitation radiation ($\lambda_{exc}$ = 980 nm) and transferring energy to $Er^{3+}$ ions, which act as acceptors. Although the energy of the excitation radiation matches the energy gap between the ground state $^4I_{15/2}$ and the excited state $^4I_{11/2}$ of $Er^{3+}$ ions, enabling potential direct absorption, the significantly higher absorption cross-section of $Yb^{3+}$ ions and the longer lifetime of their excited state $^2F_{5/2}$ compared to the $^4I_{11/2}$ state of $Er^{3+}$ favor energy transfer up-conversion mechanisms. The initial energy transfer from $Yb^{3+}$ to $Er^{3+}$ is followed by a second transfer that populates the $^4F_{7/2}$ level. This level typically undergoes nonradiative relaxation via multiphonon processes to populate the $^4S_{3/2}$ level, and subsequently, through thermalization processes (discussed later), the $^2H_{11/2}$ level. Although this is the most commonly reported mechanism in the literature, Berry et al.[37] have demonstrated that an additional pathway may exist: a subsequent energy transfer from $Yb^{3+}$ to $Er^{3+}$ can result in absorption of another excitation photon from the $^4S_{3/2}$ level, promoting electrons to the $^2D_{5/2}$ level. Nonradiative relaxation from this level leads to population of the $^2H_{9/2}$ level, resulting in emission bands at approximately 410 nm (corresponding to the $^2H_{9/2} \rightarrow ^4I_{15/2}$ transition) and around 575 nm ($^2H_{9/2} \rightarrow ^4I_{13/2}$ transition). Radiative relaxation from the $^4S_{3/2}$ and $^2H_{11/2}$ levels results in emission bands centered at 550 nm and 520 nm, respectively. These two levels are thermally coupled, meaning that as temperature increases, the relative population of the $^2H_{11/2}$ level increases at the expense of the $^4S_{3/2}$ level, following Boltzmann statistics. Because the absorption of a third photon by $Er^{3+}$ occurs from the $^4S_{3/2}$ level, thermalization to the $^2H_{11/2}$ level reduces the population available in $^4S_{3/2}$, thereby decreasing the intensity of luminescence from the $^2H_{9/2}$ level. Regarding the population of the $^4F_{9/2}$ level, two principal mechanisms are considered. The first involves multiphonon relaxation from the $^4S_{3/2}$ level. The second suggests that after the first energy transfer populates the $^4I_{11/2}$ level, it undergoes nonradiative relaxation to the $^4I_{13/2}$ level, from which the next photon is absorbed, directly populating the $^4F_{9/2}$ level. This second mechanism is more probable in host materials with high phonon energy, or when the energy transfer between $Yb^{3+}$ and $Er^{3+}$ is inefficient such as when the distance between donor and acceptor ions is large, or when the excitation power density is low. It is worth noting that both of the aforementioned population pathways for the $^4S_{3/2}$ and $^4F_{9/2}$ levels may operate concurrently, with one dominating depending on specific conditions. Radiative depopulation of the $^4F_{9/2}$ level results in a red emission band attributed to the $^4F_{9/2} \rightarrow {}^4I_{15/2}$ transition. Given that the efficiency of energy transfer between $Er^{3+}$ and $Yb^{3+}$ ions depends significantly on their average interionic distance, the spectroscopic properties of $Na_3Sc_2(PO_4)_3$:$Er^{3+}$, $Yb^{3+}$ were investigated across a series of phosphors with varying $Yb^{3+}$ concentrations. A comparison of the room-temperature up-conversion spectra for these samples consistently revealed four emission bands centered at approximately 525 nm, 550 nm, 575 nm, and 670 nm. These bands correspond to the electronic transitions described above. Notably, the presence of the $^2H_{9/2} \rightarrow {}^4I_{13/2}$ emission band in the spectra confirms the occurrence of the three-photon excitation mechanism previously outlined for this material system. Although the spectral positions of the $Er^{3+}$ emission bands remain unchanged with increasing sensitizer concentration, a noticeable enhancement in the intensity of the $Er^{3+}$ emission bands in the green region of the spectrum is observed relative to those in the red region. This effect suggests that the reduction in the average distance between $Er^{3+}$ and $Yb^{3+}$ ions, resulting from the increased $Yb^{3+}$ concentration, facilitates more efficient population of higher-lying energy levels of $Er^{3+}$ ions. An analysis of the ratio of luminescence intensity in the green spectral region to that in the red region reveals a monotonic increase from approximately 0.2 for 5 %$Yb^{3+}$ to 0.4 for 30 %$Yb^{3+}$. These variations are clearly manifested in the color of the light emitted by $Na_3Sc_2(PO_4)_3$:$Er^{3+}$, $Yb^{3+}$, shifting from orange for 5 %$Yb^{3+}$ to greenish for 30 %$Yb^{3+}$.

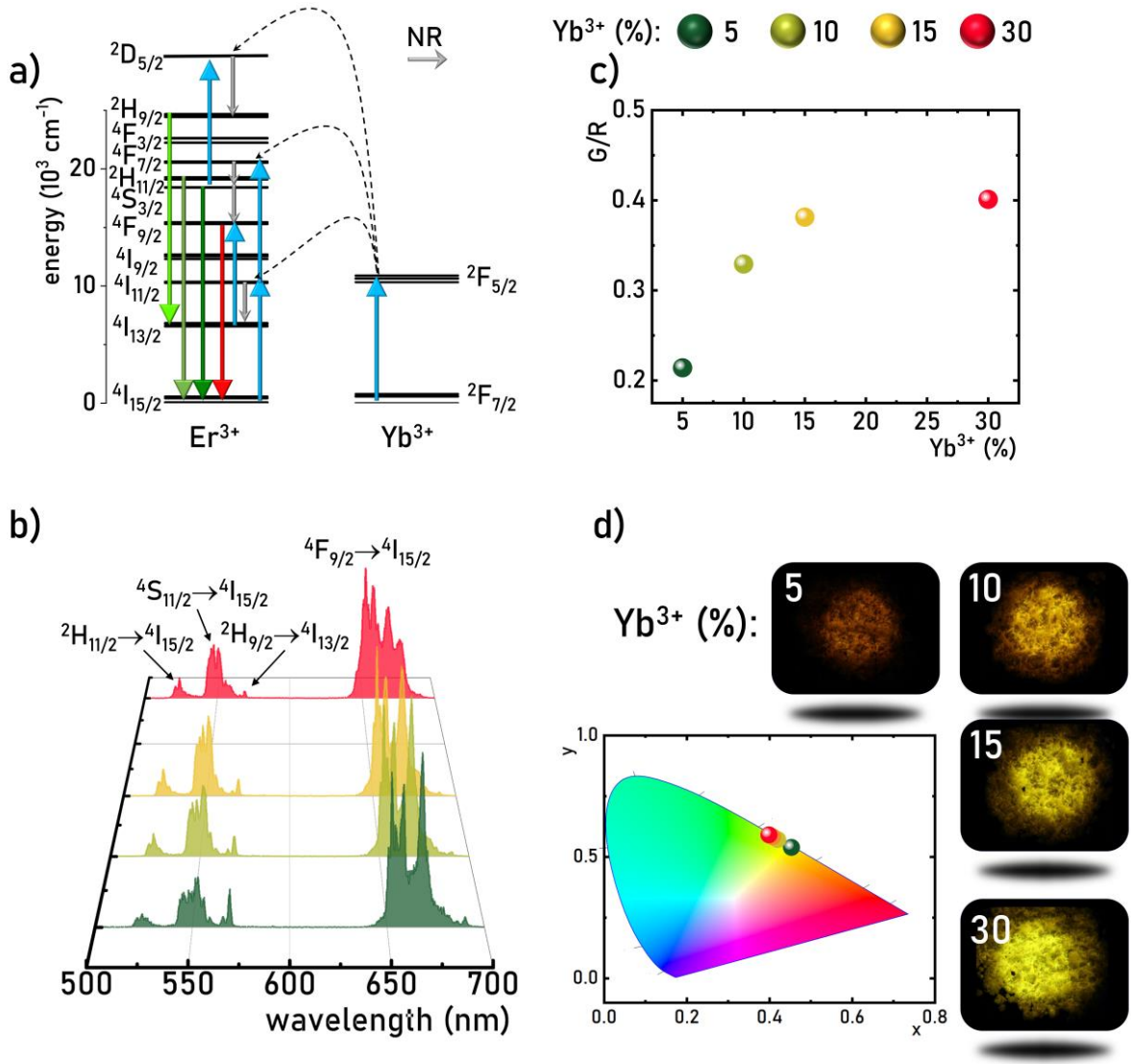

**Figure 2.** Schematic energy levels diagram of the $Yb^{3+}$-$Er^{3+}$ co-doped system -a), comparison of room-temperature emission spectra ($\lambda_{exc}$ = 980 nm) -b), green-to-red emission intensity ratio -c), and CIE chromaticity coordinates with corresponding luminescence photographs for $Na_3Sc_2(PO_4)_3$:$Er^{3+}$, $xYb^{3+}$ (x = 5, 10, 15, and 30%).

To evaluate the influence of temperature on the up-conversion luminescence behavior of $Na_3Sc_2(PO_4)_3$:$Er^{3+}$, $Yb^{3+}$, their temperature-dependent luminescence spectra were recorded over the range of 83 K to 603 K. Representative spectra (Figure 3a) for $Na_3Sc_2(PO_4)_3$:0.5%$Er^{3+}$, 5%$Yb^{3+}$ reveal that at 83 K, the emission is predominantly composed of bands corresponding to the $^4S_{3/2} \rightarrow {}^4I_{15/2}$, and $^4F_{3/2} \rightarrow {}^4I_{15/2}$ transitions, with a weak signal also observed from the

$^2H_{9/2}$ level (see also Figure S3-S5). As the temperature increases, a general decrease in the intensity of all emission bands is noted; however, the most pronounced changes occur within the green spectral region (Figure 3b). Notably, the $^2H_{9/2} \rightarrow {}^4I_{13/2}$ band shows a distinct thermal response, with its intensity increasing up to a maximum at approximately 250 K, followed by a progressive decline. Above ~450 K, the intensity of this band becomes negligible. Interestingly, the temperature range over which the $^2H_{9/2} \rightarrow {}^4I_{13/2}$ band diminishes corresponds closely with the thermal activation of the $^2H_{11/2} \rightarrow {}^4I_{15/2}$ band. Concurrently, the intensity of the $^4S_{3/2} \rightarrow {}^4I_{15/2}$ band begins to decrease, reflecting thermal coupling between the $^2H_{11/2}$ and $^4S_{3/2}$ levels. Beyond 480 K, the emission band corresponding to the $^2H_{11/2} \rightarrow {}^4I_{15/2}$ electronic transition dominates the green region of the spectrum. To further assess these variations, the integral intensities of individual $Er^{3+}$ emission bands were quantified as a function of temperature (Figure 3c). For a sample doped with 5% $Yb^{3+}$, the $^4F_{9/2} \rightarrow {}^4I_{15/2}$ band maintains stable in intensity up to ~350 K, after which it begins to decline. In contrast, the $^4S_{3/2} \rightarrow {}^4I_{15/2}$ and $^2H_{9/2} \rightarrow {}^4I_{13/2}$ bands exhibit thermal enhancement up to ~350 K, followed by a decrease in emission with further temperature increase. The most significant enhancement is observed for the $^2H_{9/2} \rightarrow {}^4I_{13/2}$ band, whose intensity at 250 K is approximately five times higher than at 83 K. The emission intensity of the band corresponding to the $^2H_{11/2} \rightarrow {}^4I_{15/2}$ transition exhibits a sharp rise in intensity up to ~300 K, after which it continues to increase, albeit at a reduced rate.

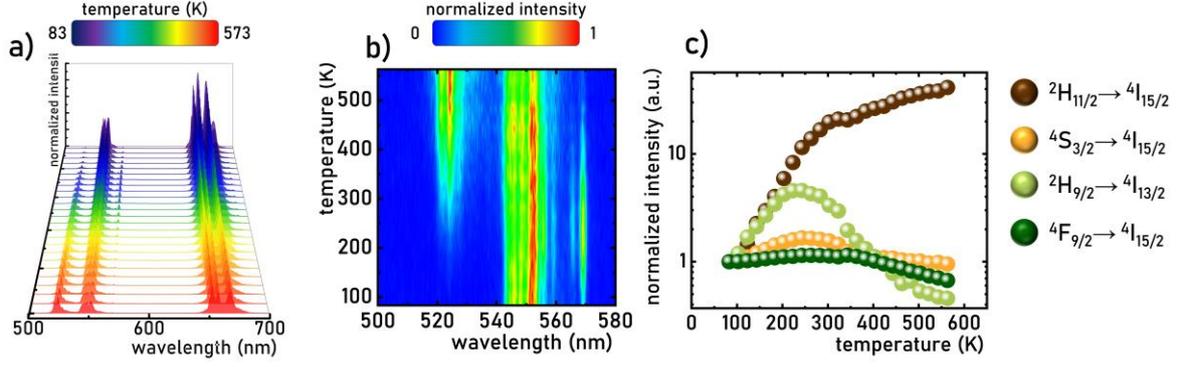

**Figure 3**. Emission spectra measured as a function of temperature - a), thermal map of normalized emission spectrum shown in the 500 nm – 580 nm spectral range – b) and the influence of the temperature on the normalized integrated emission intensities of different emission bands of $Er^{3+}$ ions -c) for $Na_3Sc_2(PO_4)_3:Er^{3+}$, 5%$Yb^{3+}$.

Power-dependent studies of up-conversion emission intensity provide critical insights into the population mechanisms of the excited states of $Eu^{3+}$ ions involved in the up-conversion process. According to the methodology proposed by Pollnau et al.[26], within the low optical excitation power density regime, the luminescence intensity ($I$) follows a power-law relationship with the excitation power density ($P$):

$$I \sim P^N \quad (1)$$

where, $N$ refers to the process order and denotes the number of photons involved in populating the emitting state. It is essential to perform such analyses under low power density conditions, as higher excitation densities can lead to saturation effects that artificially lower the observed process order. To investigate these dynamics, the dependence of up-conversion luminescence intensity on optical power density was measured for $Na_3Sc_2(PO_4)_3:0.5\%Er^{3+}$, 5%$Yb^{3+}$ and $Na_3Sc_2(PO_4)_3:0.5\%Er^{3+}$, 30%$Yb^{3+}$ (Figure 4a, see also Figure S6 and S7). The results clearly show an increase in emission intensity with rising power density. Notably, the green emission bands exhibit a steeper increase than the $^4F_{9/2} \rightarrow {}^4I_{15/2}$ emission band. For instance, the integrated intensity of the $^4S_{3/2} \rightarrow \, ^4I_{15/2}$ band for a sample containing 5% $Yb^{3+}$ yields a process order $N = 2.95$, with a decline observed beyond ~50 W·cm$^{-2}$, indicating saturation (Figure 4b). When the $Yb^{3+}$ concentration is increased to 30%, $N$ for the same band decreases to 2.3, suggesting enhanced energy transfer efficiency due to reduced interionic distances. Similarly high process order values are obtained for the $^2H_{9/2} \rightarrow \, ^4I_{13/2}$ emission band, where $N \approx 3$, confirming its three-photon excitation origin, consistent with the high energy of this excited level (Figure 4c). While the $^4S_{3/2}$ level is generally associated with two-photon excitation as reported in the literature[12,27,28,30], the present results indicates unusual three-photon process under low $Yb^{3+}$ concentrations[37]. However, for higher concentration of $Yb^{3+}$ ions, for which the interionic distance between sensitizer and activator is shortened, facilitates the $Yb^{3+} \rightarrow Er^{3+}$ energy transfer, transitioning population scheme to a conventional two-photon mechanism. In contrast, the $^4F_{9/2}$ level displays an $N$ value close to 2 across all concentrations, suggesting a consistent two-photon excitation mechanism, largely independent of sensitizer ion concentration (Figure 4d). These findings collectively elucidate the nuanced dependence of up-conversion dynamics on dopant concentration and excitation conditions.

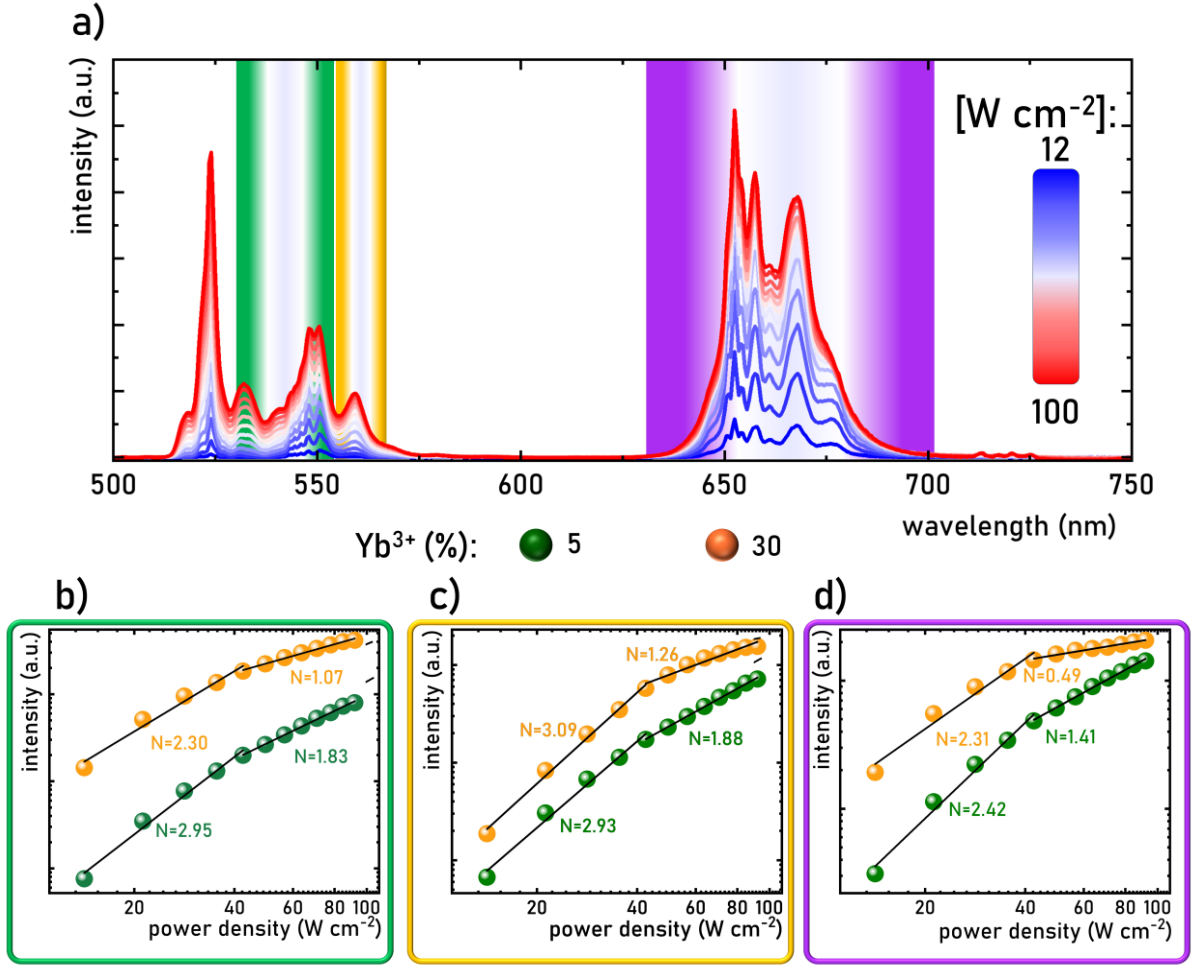

**Figure 4**. Room temperature emission spectra of $Na_3Sc_2(PO_4)_3$:$Er^{3+}$, 30%$Yb^{3+}$ measured as a function of power densities of $\lambda_{exc}$=980 nm – a); the log-log plots of emission intensities as a function of excitation density for emission bands corresponding to the $^4S_{3/2} \rightarrow {}^4I_{15/2}$ – b); $^2H_{9/2} \rightarrow {}^4I_{13/2}$ – c) and $^4F_{9/2} \rightarrow {}^4I_{15/2}$ – d) electronic transitions for $Na_3Sc_2(PO_4)_3$:$Er^{3+}$, 5%$Yb^{3+}$ and $Na_3Sc_2(PO_4)_3$:$Er^{3+}$, 30%$Yb^{3+}$.

In ratiometric luminescence thermometry, achieving high temperature sensitivity relies on selecting emission bands that exhibit opposite thermal monotonicity. The analysis of the temperature dependence of the individual $Er^{3+}$ emission bands, as presented in Figure 3c, reveals several such band pairs. Among them, the emission bands associated with the depopulation of the well-established thermally coupled $^2H_{11/2}$ and $^4S_{3/2}$ levels are particularly notable[10,11]. The thermal evolution of their emission intensities, shown in Figures 5a and 5b, demonstrates that the $^2H_{11/2}\rightarrow{}^4I_{15/2}$ band intensity increases monotonically with temperature across the entire studied range. Notably, this increase is most pronounced at low $Yb^{3+}$ concentrations and approximately a 40-fold enhancement of its emission intensity for 5% $Yb^{3+}$ was observed compared to a 20-fold increase at 30% $Yb^{3+}$. A minor deviation from the general trend of these curves is observed around 330 K, which likely corresponds to a structural phase transition discussed earlier. In contrast, the $^4S_{3/2}\rightarrow{}^4I_{15/2}$ band displays a non-monotonic trend (Figures 5b). At low $Yb^{3+}$ concentrations, its intensity initially increases with temperature up to ~250 K, followed by a gradual decrease. This thermally activated enhancement diminishes with increasing $Yb^{3+}$ concentration, and at 15% $Yb^{3+}$, a monotonic decline in emission intensity is observed throughout the temperature range. These findings indicate that at low $Yb^{3+}$ doping levels, both levels are highly responsive to temperature fluctuations. The initial enhancement of the intensity of the $^4S_{3/2}\rightarrow{}^4I_{15/2}$ emission suggests the involvement of thermally activated (likely phonon-assisted) processes in its population. At higher $Yb^{3+}$ concentrations, reduced interionic distances facilitate more efficient energy transfer, diminishing the role of such phonon-mediated processes. This interpretation is further supported by the observed reduction in the power dependence parameter $N$ for the $^4S_{3/2}\rightarrow{}^4I_{15/2}$ band at elevated $Yb^{3+}$ ions concentration. The ratio of the emission intensities from the $^2H_{11/2}\rightarrow{}^4I_{15/2}$ and $^4S_{3/2}\rightarrow{}^4I_{15/2}$ levels ($LIR_1$) is defined as:

$$LIR_1 = \frac{\int_{514nm}^{536nm}\left(^2H_{11/2}\rightarrow{}^4I_{15/2}\right)d\lambda}{\int_{536nm}^{554nm}\left(^4S_{3/2}\rightarrow{}^4I_{15/2}\right)d\lambda} \qquad (2)$$

If these levels remain thermally coupled, the ratio should follow a Boltzmann-type dependence[11] []. Indeed, as shown in Figure 5c, the natural logarithm of $LIR_1$ exhibits a linear relationship with the inverse of temperature (1/T) in the range of 0.002 to 0.010 $K^{-1}$. Remarkably, the slope of this linear relationship remains nearly constant across all $Yb^{3+}$ ions concentrations, confirming that thermal coupling is preserved and the energy separation between the two levels remains unaffected by doping concentration. This consistency implies that the observed differences in temperature dependence are attributed primarily to variations in the population mechanisms of the levels, rather than to alternative depopulation pathways or energy transfer processes. The performance of this thermometric approach is quantified by the relative sensitivity, defined as:

$$S_R = \frac{1}{LIR}\frac{\Delta LIR}{\Delta T} \cdot 100\% \qquad (3)$$

The calculated values of $S_{R1}$ are similar across all $Yb^{3+}$ concentrations, with sensitivities decreasing from approximately 2.8%·K$^{-1}$ at 200 K to 0.8%·K$^{-1}$ at 300 K, and further to 0.4%·K$^{-1}$ at 550 K (Figures 5d). These results highlight the reliability and robustness of the ratiometric approach based on $^2H_{11/2} \rightarrow {}^4I_{15/2}$ to $^4S_{3/2} \rightarrow {}^4I_{15/2}$ bands in $Na_3Sc_2(PO_4)_3$:$Er^{3+}$, $Yb^{3+}$ for temperature sensing applications over a broad temperature range and various sensitizer concentrations.

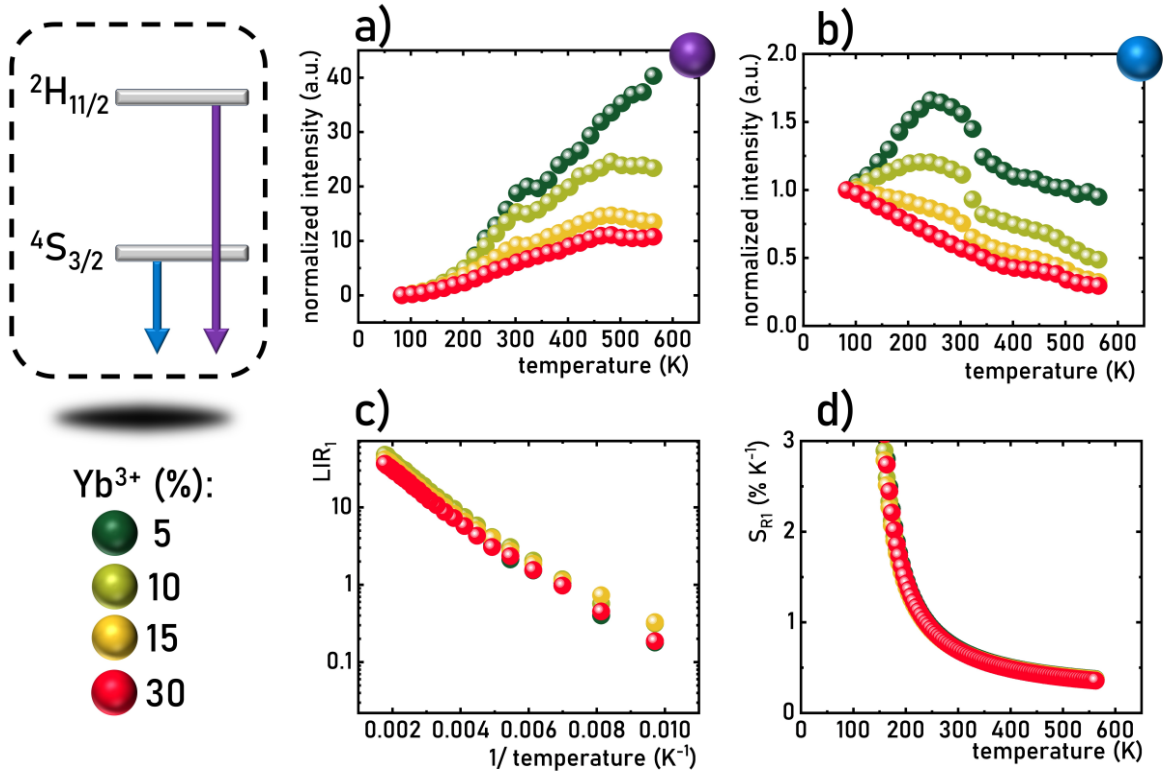

**Figure 5**. Thermal dependence of normalized integral emission intensities of emission bands corresponding to the $^2H_{11/2} \rightarrow {}^4I_{15/2}$ – a) and $^4S_{3/2} \rightarrow {}^4I_{15/2}$ – b) for different dopant concentration of $Yb^{3+}$ ions; $LIR_1$ as a function of 1/T -c) and corresponding thermal dependence of $S_{R1}$ -d).

As previously demonstrated, the intensity of the $^2H_{9/2} \rightarrow {}^4I_{13/2}$ emission band also increases with rising temperature; however, this enhancement is observed only up to approximately 250 K. Beyond this point, further temperature increases result in a gradual decline in its luminescence intensity. Similar to the behavior of the $^2H_{11/2} \rightarrow {}^4I_{15/2}$ band, the thermal enhancement of the $^2H_{9/2} \rightarrow {}^4I_{13/2}$ band decreases with increasing $Yb^{3+}$ ion concentration, suggesting that this effect is closely linked to the mechanism of level population. Nevertheless, a comparative analysis of the thermal growth rates of the $^2H_{9/2} \rightarrow {}^4I_{13/2}$ (Figure 6a) and $^4S_{3/2} \rightarrow {}^4I_{15/2}$ emission bands (Figure 6b) reveals that their intensity ratio ($LIR_2$) can be employed for ratiometric temperature sensing:

$$LIR_2 = \frac{\int_{562nm}^{572nm} \left(^2H_{9/2} \to {}^4I_{15/2}\right) d\lambda}{\int_{536nm}^{554nm} \left(^4S_{3/2} \to {}^4I_{15/2}\right) d\lambda} \quad (4)$$

As shown, $LIR_2$ exhibits a consistent thermal trend across all $Yb^{3+}$ concentrations, rising with temperature up to ~275 K, followed by a decline (Figure 6c). Since reliable temperature readout in ratiometric luminescence thermometry requires monotonic variation of $LIR$, this inflection limits the applicable thermal range of $LIR_2$-based sensing to a maximum of ~275 K. This behavior is further reflected in the thermal dependence of the relative sensitivity $S_{R2}$ (Figure 6d), which peaks at ~3% $K^{-1}$ near 100 K and gradually decreases, becoming negative above ~280 K. Although negative $S_R$ values are not suitable for quantitative sensing, they are presented here to illustrate the loss of thermal monotonicity in $LIR_2$. Notably, due to the fact that the $^2H_{9/2} \to {}^4I_{13/2}$ emission band is observed at lower temperatures than the $^2H_{11/2} \to {}^4I_{15/2}$ band, $LIR_2$-based thermometry extends the operational range of the $Na_3Sc_2(PO_4)_3$:$Er^{3+}$, $Yb^{3+}$-based luminescent thermometers down to sub-200 K temperatures, providing complementary thermometric capabilities to those based on $LIR_1$.

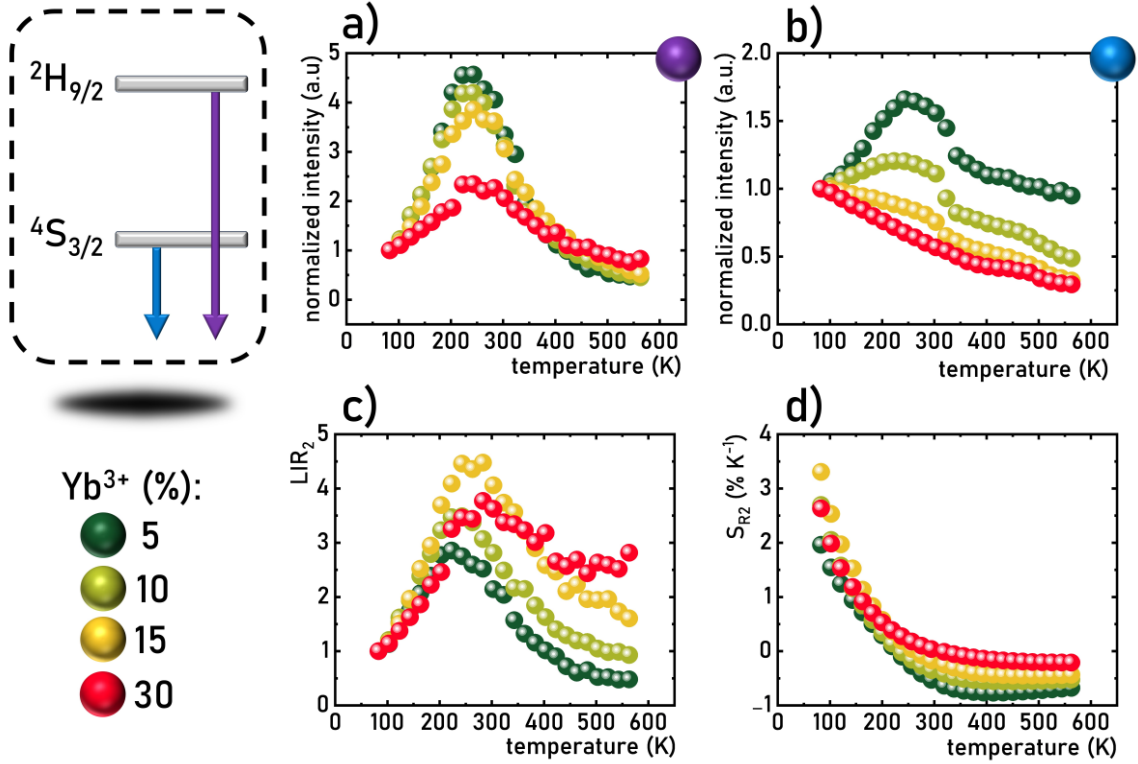

**Figure 6**. Thermal dependence of normalized integral emission intensities of emission bands corresponding to the $^2H_{9/2} \rightarrow {}^4I_{13/2}$ – a) and $^4S_{3/2} \rightarrow {}^4I_{15/2}$ – b) for different dopant concentration of $Yb^{3+}$ ions; thermal dependence of $LIR_2$ – c) and corresponding thermal dependence of $S_{R2}$ – d).

A comprehensive analysis of the temperature-dependent intensity variations of $Er^{3+}$ emission bands in $Na_3Sc_2(PO_4)_3$:0.5%$Er^{3+}$, 5%$Yb^{3+}$ reveals a pronounced difference between the thermal behavior of bands located in the green spectral region and those in the red region. This distinction is particularly relevant for practical applications, as the green emission bands are spectrally well-separated from the $^4F_{9/2} \rightarrow {}^4I_{15/2}$ band, unlike $LIR_1$ and $LIR_2$. This spectral separation facilitates more accurate temperature readings and, more importantly, enables visual temperature sensing due to the resulting changes in emission color. To investigate this possibility, the temperature dependence of the total emission intensities of the $^2H_{11/2} \rightarrow {}^4I_{15/2}$, $^4S_{3/2} \rightarrow {}^4I_{15/2}$, and $^2H_{9/2} \rightarrow {}^4I_{13/2}$ bands was analyzed (Figure 7a). All three bands exhibit thermal enhancement across the studied $Yb^{3+}$ ions concentrations, although the rate of increase diminishes with higher dopant levels. In contrast, the $^4F_{9/2} \rightarrow {}^4I_{15/2}$ band displays only a slight intensity increase up to ~400 K for 5% $Yb^{3+}$, followed by thermal quenching (Figure 7b). For higher $Yb^{3+}$ concentrations, no thermal enhancement is observed -only intensified thermal quenching.

These findings allowed the definition of a third luminescence intensity ratio ($LIR_3$) as follows:

$$LIR_3 = \frac{\int\limits_{514nm}^{580nm} \left({}^2H_{11/2} \rightarrow {}^4I_{15/2}\right) + \left({}^4S_{3/2} \rightarrow {}^4I_{15/2}\right) + \left({}^2H_{9/2} \rightarrow {}^4I_{13/2}\right) d\lambda}{\int\limits_{640nm}^{700nm} \left({}^4F_{9/2} \rightarrow {}^4I_{15/2}\right) d\lambda} \quad (5)$$

which, as shown in Figure 7c, increases by over 50-fold upon heating to 553 K. The relative sensitivity ($S_{R3}$) peaks at 1.6% $K^{-1}$ at 100 K for 10% $Yb^{3+}$, indicating strong thermometric responsiveness (Figure 7d). This substantial variation in $LIR_3$ corresponds to a noticeable shift in the emission color, confirmed by the calculated CIE 1931 chromaticity coordinates (Figure S9-S12). For example, with 5% $Yb^{3+}$, the emission color changes from orange at 83 K to greenish-yellow at 553 K, while for 30% $Yb^{3+}$, the transition spans from yellow to green over the same temperature range (Figure 7e). Moreover, analysis of the temperature evolution of the x and y chromaticity coordinates shows an inflection for low $Yb^{3+}$ concentrations, likely due to a structural phase transition, whereas 30% $Yb^{3+}$ doping yields a smooth and predictable chromatic shift (Figure 7f). These results highlight the viability of colorimetric thermometry using $Na_3Sc_2(PO_4)_3$:$Er^{3+}$, $Yb^{3+}$, offering a practical and filter-free strategy for visual temperature sensing.

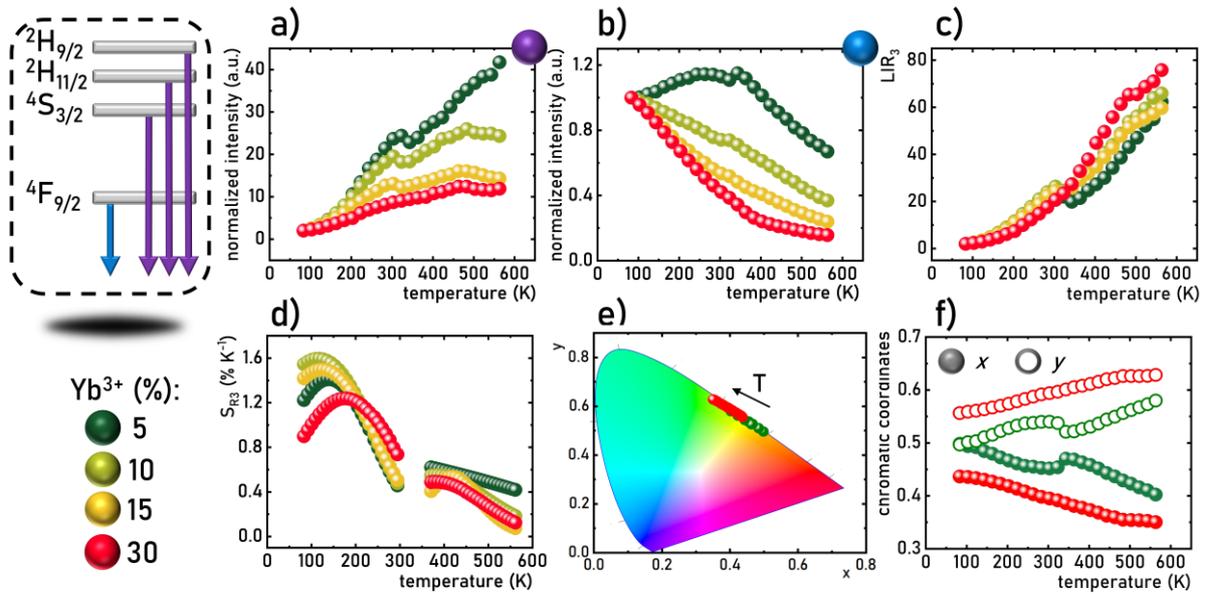

**Figure 7**. Thermal dependence of normalized integral emission intensities of green emission – total emission intensities of emission bands corresponding to the $^2H_{9/2} \rightarrow {}^4I_{13/2} + {}^2H_{11/2} \rightarrow {}^4I_{15/2} + {}^4S_{3/2} \rightarrow {}^4I_{15/2}$ – a) and red emission - $^4F_{9/2} \rightarrow {}^4I_{15/2}$ – b) for different dopant concentration of $Yb^{3+}$ ions; thermal dependence of $LIR_3$ -c) and corresponding thermal dependence of $S_{R3}$ -d); thermal dependence of CIE 1931 chromatic coordinates – e) and corresponding thermal dependence of chromatic coordinates – e).

As demonstrated, the pronounced temperature-dependent color changes in the luminescence of $Na_3Sc_2(PO_4)_3$:$Er^{3+}$, $Yb^{3+}$ can be effectively utilized for visual temperature sensing. In addition to passive temperature variations, the sample temperature can also be actively modulated by adjusting the optical power density of the excitation source. This is due to the partial conversion of absorbed excitation energy into heat, rather than up-conversion emission. Consequently, the interplay between thermally induced luminescence color shifts and optically induced heating can be harnessed to develop a visual optical power density sensor. Given that the absorption efficiency of excitation light is proportional to the concentration of sensitizer ions, the emission color change under varying optical power densities was investigated for two extreme $Yb^{3+}$ concentrations: 5% and 30%. Digital images reveal that for 5% $Yb^{3+}$, the sample emits a weak orange glow at low excitation densities, which gradually shifts to yellow and finally to a greenish-yellow hue at 104 W cm$^{-2}$ (Figure 8a, Figure S13-S16). In contrast, the 30% Yb$^{3+}$-doped sample exhibits significantly stronger luminescence across the same power density range, with a pronounced color shift from orange to bright green (Figure 8b, Figure S17-20). Additionally, surface temperature measurements confirmed the enhanced optical heating efficiency in the sample with 30% Yb$^{3+}$, reaching 534 K at 104 W cm$^{-2}$, compared to only 329 K for the 5% Yb$^{3+}$ counterpart. These findings demonstrate the potential of Na$_3$Sc$_2$(PO$_4$)$_3$:Er$^{3+}$, Yb$^{3+}$ as a dual-function material for both luminescent temperature sensing and visual optical power density mapping. The variation in the color of light emitted by Na$_3$Sc$_2$(PO$_4$)$_3$:Er$^{3+}$, Yb$^{3+}$ in response to changes in excitation power density is clearly reflected in the shift of the corresponding CIE 1931 chromaticity coordinates (Figure 8c). For the sample doped with 30% Yb$^{3+}$, these coordinates transition from values characteristic of orange emission to those indicative of green emission across the analyzed power density range. A detailed analysis of these changes reveals substantially greater chromatic shifts for the 30% Yb$^{3+}$ sample: the *x* coordinate varies from 0.56 to 0.36, and the *y* coordinate from 0.43 to 0.61 (Figure 8d). In comparison, the 5% Yb$^{3+}$ sample exhibits more modest changes, with *x* decreasing from 0.62 to 0.52 and *y* increasing from 0.38 to 0.47. The relative sensitivity values derived from chromatic coordinate variations, calculated analogously to thermal sensitivity but using power density (p) instead of temperature (T), demonstrate that for Na$_3$Sc$_2$(PO$_4$)$_3$:Er$^{3+}$, 30%Yb$^{3+}$, maximum sensitivities reach $S_{Rx}$ = 1.0 % W$^{-1}$ cm$^2$ and $S_{Ry}$ = 0.9 % W$^{-1}$ cm$^2$ at 15 W cm$^{-2}$ (Figure 8e). These values decrease with increasing power density to 0.14 % W$^{-1}$ cm$^2$ and 0.27 % W$^{-1}$ cm$^2$, respectively, at 90 W cm$^{-2}$. In contrast, nearly two-fold lower sensitivities are observed for the 5% Yb$^{3+}$ sample. These findings underscore that higher Yb$^{3+}$ ion concentrations, which lead to more efficient absorption of excitation light and enhanced optically induced heating, are advantageous for the development of luminescent optical power density sensors.

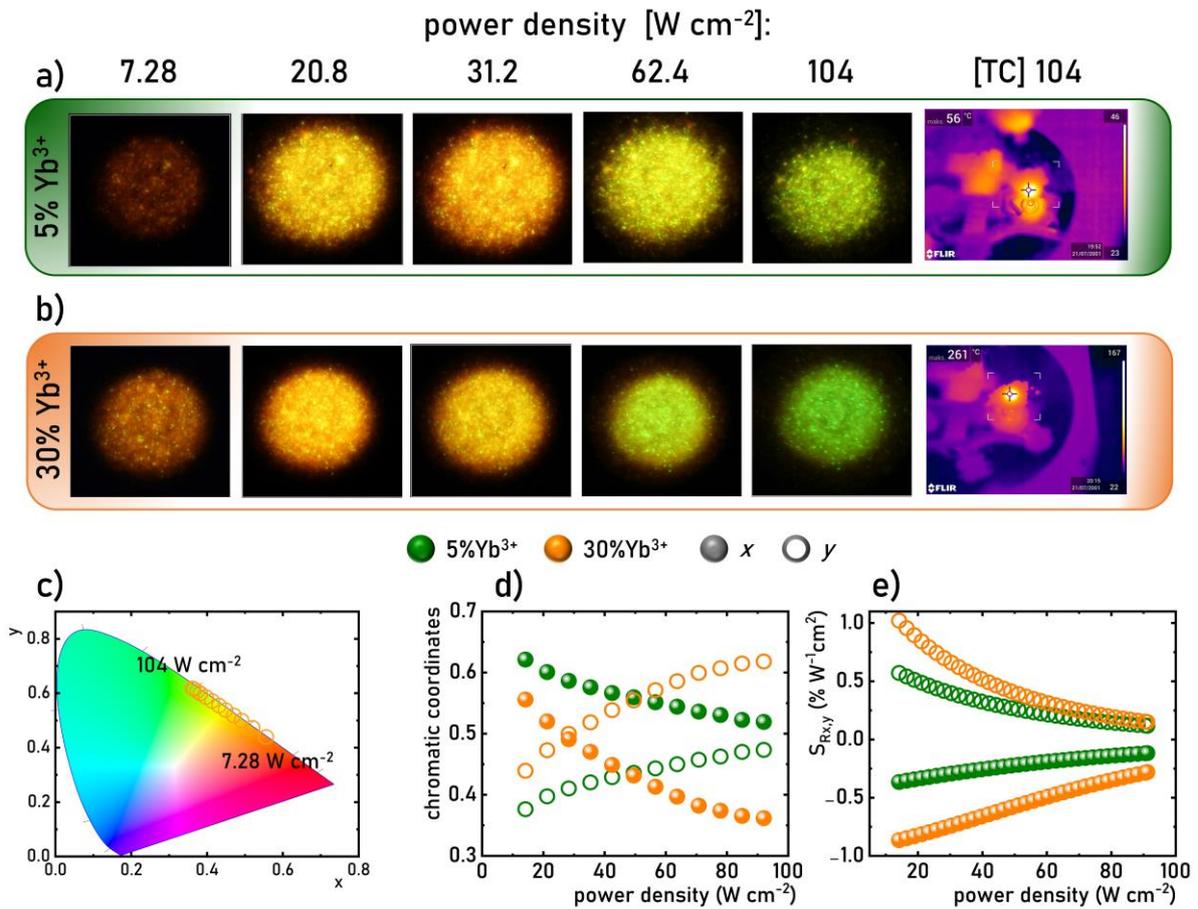

**Figure 8**. Photos of Na$_3$Sc$_2$(PO$_4$)$_3$:0.5%Er$^{3+}$, 5%Yb$^{3+}$ - a) and Na$_3$Sc$_2$(PO$_4$)$_3$:0.5%Er$^{3+}$, 30%Yb$^{3+}$ - b) captured under different excitation densities of $\lambda_{exc}$= 980 nm, and the picture from the thermovision camera; the influence of the power density on the CIE 1931 chromatic coordinates for 30%Yb$^{3+}$ - c); the influence of the excitation density on the x and y chromatic coordinates for 5%Yb$^{3+}$ and 30%Yb$^{3+}$ -d) and corresponding relative sensitivities – e).

As demonstrated above, variations in the optical power density used to excite the up-conversion luminescence of Na$_3$Sc$_2$(PO$_4$)$_3$:Er$^{3+}$, Yb$^{3+}$ result in noticeable changes in the color of the emitted light. However, performing colorimetric analysis based on CIE 1931 chromaticity coordinates requires full spectral acquisition and extensive post-processing, which is time-consuming and impractical, particularly for two-dimensional imaging applications. A more straightforward, cost-effective, and rapid alternative involves capturing luminescence images and analyzing the intensity values recorded in the red (*R*), green (*G*), and blue (*B*) channels of a standard digital camera. This method enables the generation of spatial distribution maps for each color channel, and by calculating the *G/R* intensity ratio, a corresponding map of the optical power density can be obtained. While this image-based approach has been utilized in luminescent thermometry[38,39] and thermal history sensing[40], to the best of our knowledge, it has not yet been applied to optical power density mapping. In the case of $Na_3Sc_2(PO_4)_3$:$Er^{3+}$, $Yb^{3+}$, this strategy is particularly advantageous because the relevant emission bands are spectrally resolved by the *R* channel ($^4F_{9/2} \rightarrow {}^4I_{15/2}$) and *G* channel ($^2H_{9/2} \rightarrow {}^4I_{13/2}$, $^4S_{3/2} \rightarrow {}^4I_{15/2}$, $^2H_{11/2} \rightarrow {}^4I_{15/2}$). To implement this method, a series of luminescence images of $Na_3Sc_2(PO_4)_3$:$Er^{3+}$, 30%$Yb^{3+}$ were acquired under varying optical power densities (Figure 9a). From each image, intensity maps for the *G* and *R* channels were extracted and used to compute *G/R* ratio maps. These data were used to construct a calibration curve relating *G/R* to optical power density, which followed an exponential trend up to 130 W cm$^{-2}$ (Figure 9b):

$$P(G/R) = 1.47 \cdot \exp\left(\frac{G/R}{14.176}\right) + 9.60 \qquad (6)$$

To validate the effectiveness of this approach for spatially resolved measurements, a layer of $Na_3Sc_2(PO_4)_3$:$Er^{3+}$, 30%$Yb^{3+}$ was illuminated with a 980 nm quasi-Gaussian laser beam (maximum intensity: 110 W cm$^{-2}$), and luminescence images were captured. The resulting *G/R* ratio map revealed a clear radial gradient, with a peak intensity at the center (~1.5 mm in diameter), decreasing outward (Figure 9c). Using the calibration curve, the *G/R* ratio map was converted into a spatially resolved optical power density map (Figure 9d). The obtained power distribution closely resembles the expected Gaussian profile, although minor deviations observed in the cross-sectional intensity profiles (Figure 9e) reflect the inhomogeneity of the excitation beam itself. In conclusion, the results confirm the strong application potential of $Na_3Sc_2(PO_4)_3$:$Er^{3+}$, 30%$Yb^{3+}$ as a luminescent material for real-time, visual two-dimensional imaging of optical power density distributions using a standard digital camera without the need for optical filters.

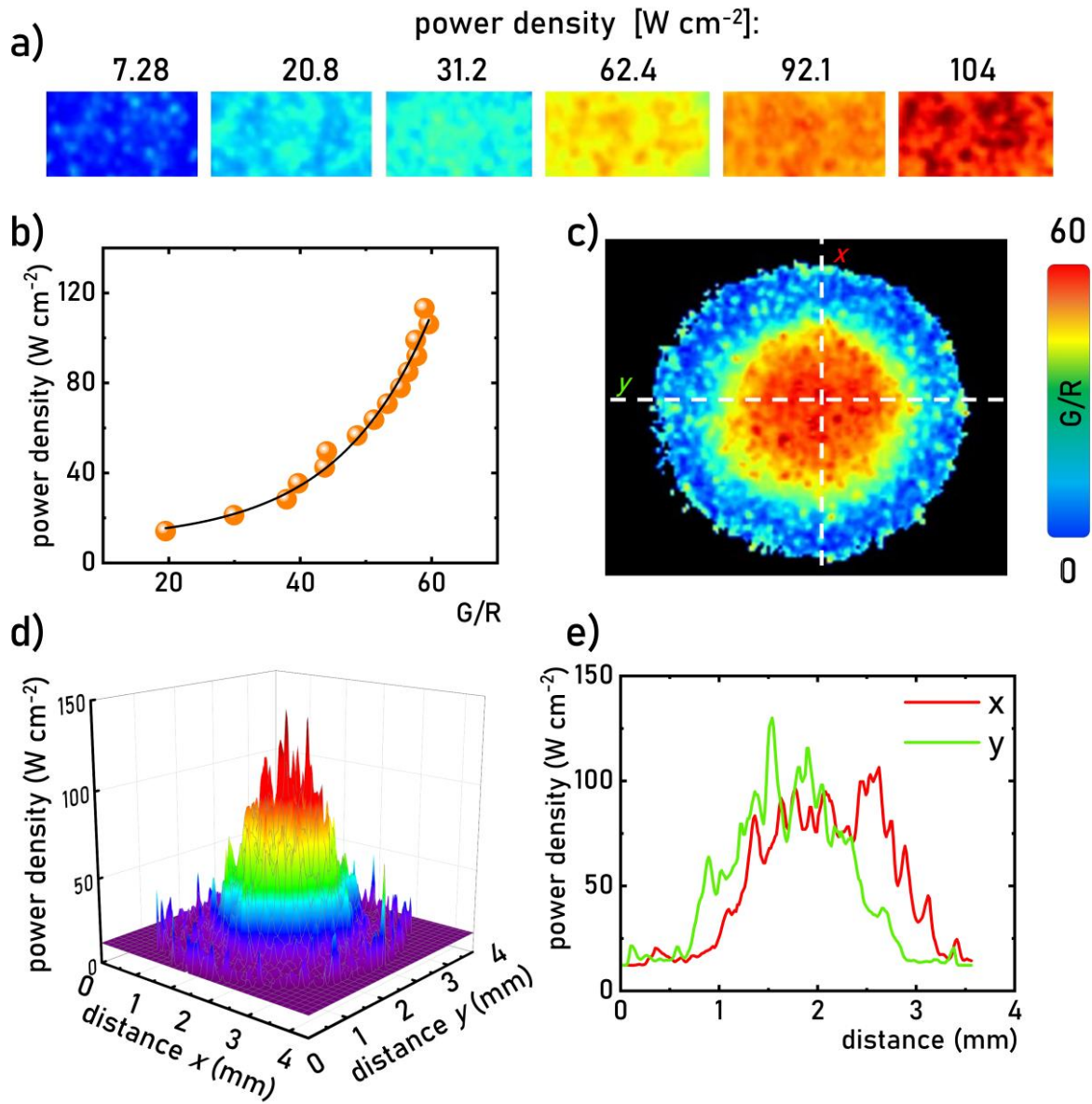

**Figure 9**. The G/R maps for $Na_3Sc_2(PO_4)_3$:$Er^{3+}$, 30%$Yb^{3+}$ determined from photos captured at different excitation densities – a); the calibration curve of power density vs G/R ratio – b); the 2D G/R maps of the $Na_3Sc_2(PO_4)_3$:$Er^{3+}$, 30%$Yb^{3+}$ luminescence obtained upon $\lambda_{exc}$=980 nm of 120 W cm$^{-2}$ excitation density - c); and corresponding 3D distribution of power density – d); and power density cross section profiles along x and y lines marked in figure 9c – e).

Several studies in the literature report the use of luminescence for sensing optical power density, primarily based on phosphors doped with $Cr^{3+}$ [41–43] or $Er^{3+}$, $Yb^{3+}$ [44,45] ions. In both cases, the sensing mechanism relies on changes in the emission intensity ratio of two thermally coupled energy levels, which are influenced by optically induced heating of the material. However, in all cases reported to date, the parameter employed for sensing is the luminescence intensity ratio, which complicates two-dimensional imaging of the optical power density distribution. For previously described sensors based on $Er^{3+}$ emission[44,45], the close spectral proximity of the bands used in sensing prevents a distinct change in the color of the emitted light. In this context, the capabilities offered by $Na_3Sc_2(PO_4)_3$:$Er^{3+}$, 30%$Yb^{3+}$ as the first visual luminescent optical power density sensor present a highly attractive alternative from an application standpoint.

**Conclusions**

In summary, this study investigated the influence of temperature on the up-conversion emission behavior of $Na_3Sc_2(PO_4)_3$:$Er^{3+}$, $Yb^{3+}$. The analysis revealed that, in addition to the characteristic emission bands of $Er^{3+}$ ions associated with the $^2H_{11/2} \rightarrow {}^4I_{15/2}$, $^4S_{3/2} \rightarrow {}^4I_{15/2}$, and $^4F_{9/2} \rightarrow {}^4I_{15/2}$ electronic transitions, an additional emission band around 575 nm was observed, which corresponds to the $^2H_{9/2} \rightarrow {}^4I_{13/2}$ transition. Significant temperature-dependent changes in the spectroscopic properties of $Na_3Sc_2(PO_4)_3$:$Er^{3+}$, $Yb^{3+}$ enabled the implementation of three distinct temperature-sensing modes based on various luminescence intensity ratios: (i) $^2H_{11/2} \rightarrow {}^4I_{15/2}$ and $^4S_{3/2} \rightarrow {}^4I_{15/2}$; (ii) $^2H_{9/2} \rightarrow {}^4I_{13/2}$ and $^4S_{3/2} \rightarrow {}^4I_{15/2}$; and (iii) green-to-red emission intensity ratio. As demonstrated, in the first mode, the concentration of $Yb^{3+}$ ions had negligible influence on the relative sensitivity, with the maximum $S_{R1max}$=2.8%·$K^{-1}$ observed at 200 K. Since the $^2H_{11/2} \rightarrow {}^4I_{15/2}$ band becomes thermally activated above 200 K, the operational temperature range for this mode spans from 200 to 553 K. The $^2H_{9/2} \rightarrow {}^4I_{13/2}$ band, on the other hand, is already detectable at 83 K, and its opposite thermal monotonicity compared to the $^4S_{3/2}$ → $^4I_{15/2}$ band allowed for the development of an alternative temperature-sensing approach. This mode yielded a maximum sensitivity of $S_{R2max}$ = 3% K$^{-1}$ at 100 K. As this mechanism remains functional below 200 K, combining both $LIR_1$ and $LIR_2$ approaches extends the overall thermal operating range of the Na$_3$Sc$_2$(PO$_4$)$_3$:Er$^{3+}$, Yb$^{3+}$-based thermometer. Moreover, a thermally induced variation in the red-to-green emission intensity ratio was observed, contributing to a noticeable color change in the emitted light from Na$_3$Sc$_2$(PO$_4$)$_3$:Er$^{3+}$, Yb$^{3+}$. This phenomenon, when combined with the efficient optical heating of Na$_3$Sc$_2$(PO$_4$)$_3$:Er$^{3+}$, Yb$^{3+}$ by the excitation beam, enabled the development of the first visual optical power density sensor.

Given that the temperature rise due to optical absorption is proportional to the sensitizer ion concentration, a visible color shift of the emitted light was recorded from orange at 7.28 W cm$^{-2}$ to green at 104 W cm$^{-2}$. Maximum relative sensitivities based on the chromaticity coordinates were $S_{Rx}$ = 1.0 % W$^{-1}$ cm$^2$ and $S_{Ry}$ = 0.9 % W$^{-1}$ cm$^2$ at 15 W cm$^{-2}$, respectively. To further simplify the optical power density readout, the ratio of intensity distribution maps recorded in the green and red channels of a standard camera was analyzed. This method enabled fast, low-cost, and two-dimensional imaging of optical power density distribution, validated by the analysis of the excitation beam profile. Altogether, the presented approach demonstrates strong potential for advancing research in the field of luminescence-based visual imaging of optical power density.

**Acknowledgements**
This work was supported by the National Science Center (NCN) Poland under project no. UMO- 2020/37/B/ST5/00164. Maja Szymczak gratefully acknowledges the support of the Foundation for Polish Science through the START program.

[45] J. C. Martins, A. Skripka, C. D. S. Brites, A. Benayas, R. A. S. Ferreira, F. Vetrone, L. D. Carlos, *Frontiers in Photonics* **2022**, *3*, 1037473.